\documentstyle[prd,preprint,aps,tighten]{revtex}

\begin{document}

\draft
\preprint{
\begin{tabular}{r}
JHU--TIPAC 96026 \\
DFTT 72/96
\end{tabular}
}

\title{Gravitational effects on the neutrino oscillation}

\author{
N. Fornengo$^{\rm a,b}$
\footnote{E--mail: fornengo@jhup.pha.jhu.edu},
C. Giunti$^{\rm b,c}$
\footnote{E--mail: giunti@to.infn.it},
C.W. Kim$^{\rm a}$
\footnote{E--mail: kim@eta.pha.jhu.edu}
and
J. Song$^{\rm a}$
\footnote{E--mail: jhsong@eta.pha.jhu.edu}
\\ \mbox{ }}

\address{
\begin{tabular}{c}
$^{\mbox{\rm a}}$ Department of Physics and Astronomy,
The Johns Hopkins University, \\ Baltimore, Maryland 21218, USA.
\\
$^{\mbox{\rm b}}$
INFN, Sezione di Torino,
via P. Giuria 1, 10125 Torino, Italy
\\
$^{\mbox{\rm c}}$
Dipartimento di Fisica Teorica, Universit\`a di Torino,
via P. Giuria 1, 10125 Torino, Italy
\end{tabular}
}
\date{November 5, 1996}
\maketitle
\begin{abstract}
The propagation of neutrinos in a gravitational field is studied.
A method of calculating a covariant quantum--mechanical phase in
a curved space--time is presented. The result is used to
calculate gravitational effects on the neutrino oscillation in the
presence of a gravitational field. We restrict
our discussion to the case of the Schwarzschild metric.
Specifically,
the cases of the radial propagation and the non--radial propagation
are considered. A possible application to gravitational lensing 
of neutrinos is also suggested.
\end{abstract}
 
\section{Introduction}

Neutrino oscillations in a flat space--time  
have been extensively studied in the past
by using both plane  waves \cite{plane} and wave packets 
\cite{packet} to represent the emitted  neutrinos. In  
particular, it has  been shown  
\cite{packet} that the standard treatment of  neutrino oscillations 
in  the plane wave approximation is valid only
for  extremely  relativistic  neutrinos,  
whereas for a  general  case, the wave
packet  treatment is essential.  In this paper,  
we discuss how  the results in a
flat space--time are  modified in a curved  
space--time. That is, we calculate the
quantum  mechanical  phase of  neutrinos that are  
produced and  propagate in a
gravitational field. Our  derivation of
neutrino  oscillation  formula in  a  
gravitational  field will be  based on the
covariant form  of the quantum  phase that  
arises due to the  assumed mixing of
massive neutrinos\cite{wudka}. First we consider the 
case of neutrinos that are emitted and
propagate  in  a  radial   trajectory in  
the   Schwarzschild   metric. Such a
gravitational effect can, in  principle, 
modify the standard vacuum oscillation
formula for the solar and  supernova 
neutrinos.  Although the size of the effect
is far  beyond the  current  experimental  detectability, 
in  particular for the solar  neutrinos, it may  
certainly be of  interest for the  neutrinos from very
massive sources. ( It is  well known that 
gravitational influence on the MSW  effect for  the solar  
neutrinos is  significant if  the equivalence principle is  
violated \cite{msw}.) In our  derivation we have not  assumed 
the weak field approximation. We then compare our results
with the  previous  results in  the case of  the radial  
trajectory  obtained by \cite{AB,GL,BHM}  with   clarifying  
remarks  on the   differences in  the  results and
interpretations.   As a  further    application, we  
also  consider  the case of non--radial propagation in  a 
gravitational field.  Finally, we discuss possible gravitational  
lensing effects on the  neutrino  oscillations, for which 
it is necessary to resort to the weak field approximation.
It is to be noted that in  the  last two  cases,  due to  the  
angular  spread of neutrinos with different masses in the
presence of the gravitational field,
a proper way  to treat the neutrinos is to resort to the wave 
packet formalism. (However, a complete, covariant description of neutrino 
propagation in a gravitational field in terms of wave packets is 
beyond the scope of this paper. This issue will be addressed elsewhere
\cite{noi}.) In order to discuss the problem in a transparent way, 
and in order to compare with the
previous analyses, we restrict ourselves to the discussion of
relativistic neutrinos, where a plane wave analysis can be
employed.

The plan of the paper is as follows. In Section II, we
briefly review the standard treatment of neutrino oscillations
in a flat space--time using the plane wave formalism.
In Section III we extend the plane
wave analysis to the case in the presence of a gravitational 
field. For definiteness, we discuss the neutrino oscillations
in a field described by the Schwarzschild metric.
Specifically
the cases of radial and non--radial propagation of neutrinos
are discussed. 
In the last part of Section III, we suggest the possibility of 
gravitational lensing of neutrinos and evaluate the 
resulting flavor--changing 
oscillation probability.

\section{Neutrino propagation in a flat space--time}
\label{sec:flat}

Let us consider a neutrino produced at a space--time point
$A(t_A, \vec x_A)$. Since it is produced by a weak interaction
process, it emerges as a flavor eigenstate $|\nu_\alpha\rangle$,
which is a superposition of the mass eigenstates $|\nu_k\rangle$, i.e.
\begin{equation}
|\nu_\alpha\rangle = \sum_k U^*_{\alpha k} |\nu_k\rangle\; ,
\end{equation}
where $U$ is the unitary mixing matrix of the neutrino fields.
What actually propagates are the
mass eigenstates, whose energy and momentum are
$E_k$ and $\vec p_k$, respectively, and they are related by the
mass--shell condition as
\begin{equation}
E_k^2 = \vec p_k^{\;2} + m_k^2\;.
\end{equation}
Both $E_k$ and $\vec p_k$ are determined by the
energy and momentum conservation at the production point 
$A$ and, in general, they are different for different
mass eigenstates. In a flat space--time, the propagation of the
state $|\nu_k\rangle$ is described by a plane wave
\begin{equation}
|\nu_k (t,\vec x)\rangle = \exp (- i \Phi_k) |\nu_k\rangle\; ,
\end{equation}
where
\begin{equation}
\Phi_k = E_k \, t - \vec p_k \cdot \vec x\; .
\end{equation}

Neutrino oscillations  take place due to the fact that different
states $|\nu_k\rangle$ propagate differently because they have 
different energies, momenta and masses. When they arrive at
a detector located at a space--time point
$B (t_B, \vec x_B)$ which detects flavor eigenstates via 
a weak interaction process, they have developed
a relative shift in their phases. In order for the oscillation to
occur and to be observed, some requirements must be met.
First, in addition to the standard assumption
of mixing of massive neutrinos,
the mass eigenstates must be produced coherently. 
This implies that the interference is possible only among mass 
eigenstates produced in the same process, because neutrinos 
produced by different processes have, in general, random relative 
phases in their wave functions, which destroy the coherence.
Secondly, the states have to be detected at the same time $t_B$ and
at the same place $\vec x_B$. 

Under these circumstances, the interference can take place and the
oscillation phenomenon arises. The probability that the neutrino 
produced as $|\nu_e\rangle$ is detected as
$|\nu_\mu\rangle$ is, therefore, (in the case of two generations, where
$U$ is parameterized as a function of
the mixing angle $\theta$ in the usual way) \cite{Kim}
\begin{equation}
{\cal P}(\nu_e\rightarrow \nu_\mu) = 
|\langle \nu_\mu |\nu_e(t_B,\vec x_B)\rangle|^2 =
\sin^2(2\theta) \sin^2\left(\frac{\Phi_{12}}{2} \right)
\label{prob}\; ,
\end{equation}
where $\Phi_{12} = \Phi_1 - \Phi_2$ and $\Phi_k$ ($k=1,2$) 
are the phases 
\begin{equation}
\Phi_k = E_k (t_B - t_A) - \vec p_k\cdot (\vec x_B - \vec x_A) =
E_k \int_{t_A}^{t_B} dt - \vec p_k\cdot 
\int_{\vec x_A}^{\vec x_B} d\vec x
\label{phase0}\; ,
\end{equation}
acquired by the mass eigenstates.

The expression for the phase $\Phi_k$ 
in Eq.(\ref{phase0}) can be written
in a covariant form, which is suitable for the subsequent application
in a curved space--time, as \cite{sto}
\begin{equation}
\Phi_k = \int_A^B p^{(k)}_\mu d x^\mu\; ,
\label{covphi}
\end{equation}
where
\begin{equation}
p^{(k)}_\mu = m_k g_{\mu\nu} \frac{d x^\mu}{ds}\; ,
\label{canon}
\end{equation}
is the canonical conjugate momentum to the coordinates $x^\mu$ and
$g_{\mu\nu}$ and $ds$ are the metric tensor and the line element,
respectively.
This covariant phase in Eq.(\ref{covphi}) was first
discussed by Stodolsky \cite{sto}, 
and has been used in \cite{AB,GL,BHM} 
to calculate the neutrino oscillation phase difference.

Equation (\ref{prob}) represents the oscillation probability 
for a neutrino produced at the space--time point 
$A(t_A, \vec x_A)$ and detected at a given space--time position 
$B(t_B, \vec x_B)$.  In actual experiments,
however, the time difference $(t_B - t_A)$ is not measured,
whereas the relative position $|\vec x_B - \vec x_A|$ 
of the source and the
detector is known. In the plane wave formalism, this can be taken
care of consistently only for relativistic neutrinos 
by replacing $(t_B-t_A)$ with \cite{BP}
\begin{equation}
(t_B - t_A) \simeq |\vec x_B - \vec x_A|\; ,
\label{teqx}
\end{equation}
thus the time difference does not appear in the formula for 
the oscillation probability.
In this approximation, the phase shift of Eq.(\ref{phase0})
becomes
\begin{equation}
\Phi_k = (E_k - |\vec p_k|) |\vec x_B - \vec x_A|\; .
\label{phiflat}
\end{equation}
Applying the relativistic expansion $ m_k \ll E_k $, we can approximate,
to the first order,
\begin{equation}
E_k
\simeq
E_0
+
{\mathrm{O}}
\left(
\frac{ m_k^2 }{ 2 \, E_0 }
\right)
\;,
\label{rel}
\end{equation}
where $E_0$ is the energy for a massless neutrino. Therefore,
we have
\begin{equation}
E_k - |\vec p_k| = E_k - \sqrt{E_k^2 - m_k^2} \simeq 
\frac{m_k^2}{2E_0}\; ,
\end{equation}
which leads to the standard result for the phase
\begin{equation}
\Phi_k \simeq \frac{m_k^2}{2E_0} |\vec x_B - \vec x_A|\;.
\label{phase1}
\end{equation}
The phase difference responsible for the oscillation can be given
by Eq.(\ref{phase1}) as
\begin{equation}
\Phi_{kj} \simeq \frac{\Delta m_{kj}^2}{2E_0} |\vec x_B - \vec x_A|\;,
\label{phaseshift}
\end{equation}
where $\Delta m_{kj}^2 = m_k^2 - m_j^2$.

For more general situations, where some or all of the states
$\nu_k$ are non--relativistic, the above discussion cannot be
applied, and a wave packet analysis is required \cite{packet}. 
In this case, the relation in Eq.(\ref{teqx}) 
is no longer valid, and moreover 
the problem of the coherence of the different states at the detection
position has to be taken into account. However, for relativistic 
neutrinos, the wave packet formalism shows that the approximation of
Eq.(\ref{teqx}) is indeed appropriate, and the oscillation probability
${\cal P}(\nu_e\rightarrow \nu_\mu)$ 
has the form of Eq.(\ref{prob}), where
the phase shift $\Phi_{kj}$ is given by Eq.(\ref{phaseshift}).

\section{Neutrino propagation in the Schwarzschild metric}
\label{sec:curved}

Let us now turn to the discussion of the propagation of neutrinos
in a gravitational field. 
For the sake of definiteness, and also because it may represent a situation
of possible physical interest, we will discuss the propagation in
a gravitational field of a non--rotating spherically symmetric object,
which is described by the Schwarzschild metric. The situation under
consideration can be described by the line element in the 
coordinate frame \{$t,r,\vartheta,\phi$\} as
\begin{equation}
ds^2 =
B(r) dt^2 -
B(r)^{-1} dr^2 -
r^2 d\vartheta^2 - r^2\sin^2\vartheta d\phi^2 \;,
\label{eq:s_metric}
\end{equation}
where
\begin{equation}
B(r) = \left( 1-\frac{2GM}{r}\right)\; ,
\end{equation}
and $G$ is the Newtonian constant 
and $M$ denotes the mass of the source
of the gravitational field. 
Since the gravitational field is isotropic, the classical orbit 
may be confined to a plane. Hence, we can choose it to be on 
the equatorial plane $\vartheta= \pi/2$, and we have $d\vartheta=0$.

The relevant components of the canonical momentum $p^{(k)}_\mu$ 
of Eq.(\ref{canon}) are

\begin{eqnarray}
p_t^{(k)} &=& m_k B(r) \frac{dt}{ds} 
\label{eq:E}\; ,
\\
p_r^{(k)} &=& -\,m_k\, B(r)^{-1} \frac{dr}{ds} 
\label{eq:p}\; ,
\\
p^{(k)}_\phi &=& \,-\,m_k r^2 \frac{d\phi}{ds} 
\label{eq:J}\; .
\end{eqnarray}
and they are related to each other and to the mass $m_k$
by the mass--shell relation
\begin{eqnarray}
m^2_k &=& g^{\mu\nu} p^{(k)}_{\mu} p^{(k)}_\nu
\label{shell}
\\ \nonumber
&=& 
\frac{1}{B(r)} (p_t^{(k)})^2 - B(r) (p_r^{(k)})^2 -
\frac{(p_\phi^{(k)})^2}{r^2}\; .
\end{eqnarray}

The fact that the metric tensor components 
do not depend on the coordinates 
$t$ and $\phi$ ensures that their canonical momenta $p_t^{(k)}$ and
$p_\phi^{(k)}$ are constant along the trajectory. We define the
constant of motion to be $E_k\equiv p_t^{(k)}$ 
and $J_k\equiv -p_\phi^{(k)}$. 
They represent the energy and the angular momentum which an observer, located
at $r=\infty$, sees for the mass eigenstate $\nu_k$. 
They differ from the energy and the angular momentum measured by an observer 
at a position $r_B$ or those at production point $r_A$. 
The correct way to define the energies which are
actually involved in a realistic situation is not, in general, unique.
For example, for a neutrino produced in the almost stationary shock wave
of a supernova, a local static reference frame for the production point
$r_A$ seems appropriate. On the
contrary, for neutrinos produced in the accretion disk around a black
hole, a free--falling orbiting system seems proper.
Similar arguments apply to the detection of neutrinos. 
For example, in the case of solar neutrinos, 
the detectors are in the free--falling frame.
The general
situation can be rather complicated and every case must be carefully
dealt with. In our discussion we will choose
the local reference frame.
The local energy, defined as the energy measured by an observer
at rest at a position $r$, is related to $E_k$
from the transformation law which relates the local reference frame
$\{ x^{\hat{\alpha}} \} 
= \{\hat{t},\hat{r},\hat{\varphi},\hat{\theta} \}$
to the frame $\{ x^{\mu}\} = \{t,r,\varphi,\theta \}$
\cite{MTW}
\begin{equation}
x^{\hat{\alpha}} = {L^{\hat{\alpha}}}_{\mu}\, x^\mu
\;\;\;\; ; \;\;\;
g_{\mu\nu} =
{L^{\hat{\alpha}}}_{\mu}
\,
{L^{\hat{\beta}}}_{\nu}
\,
\eta_{\hat{\alpha}\hat{\beta}}
\;,
\end{equation}
where
$ {L^{\hat{\alpha}}}_{\mu} $
are the coefficients of the transformation between the two bases:
\begin{equation}
{L^{\hat{t}}}_{t}
=
\sqrt{|g_{tt}|}
\;,
\quad
{L^{\hat{r}}}_{r}
=
\sqrt{|g_{rr}|}
\;,
\quad
{L^{\hat{\vartheta}}}_{\vartheta}
=
\sqrt{|g_{\vartheta\vartheta}|}
\;,
\quad
{L^{\hat{\varphi}}}_{\varphi}
=
\sqrt{|g_{\varphi\varphi}|}
\;,
\quad
\mbox{others}=0
\;.
\end{equation}
Therefore, the local energy is
\begin{eqnarray}
E_k^{(loc)} (r) &=& |g_{tt}|^{-1/2} E_k = B(r)^{-1/2} E_k\;.
\label{eloc}
\end{eqnarray}

In order to obtain the
neutrino oscillation probability in
a gravitational field,
we will calculate the
interference of the wave functions
of different mass eigenstates
created at a space--time point $A$
and detected at a space--time point $B$.
In the plane wave approximation, the phase of each
mass eigenstate $\nu_k$ is defined by the covariant
expression in Eq.(\ref{covphi}) and the interference
of the $k^{\mathrm{th}}$
and
$j^{\mathrm{th}}$
mass eigenstates
is given by the
phase difference
\begin{equation}
\Phi_{kj} = \int_A^B
\left( p^{(k)}_\mu - p^{(j)}_\mu \right)
d x^\mu
=
\Phi_{k}
-
\Phi_{j}
\;.
\label{101}
\end{equation}
Here the integration must be made on a definite space--time trajectory
from $A$ to $B$. Following the standard treatment
of the oscillations of the \emph{relativistic} neutrinos
in a flat space--time, as discussed in Section \ref{sec:flat},
we will calculate the interference phase in Eq.(\ref{101})
along the light--ray trajectory from $A$ to $B$. This corresponds
to the approximation in Eq.(\ref{teqx}) for the flat space--time case.
We emphasize that the phases in Eq.(\ref{101})
are \emph{not} the phases on the classical trajectory
of the mass eigenstates \cite{sto}
but the phases calculated on the light--ray trajectory.
We will see that for relativistic neutrinos
the result for the phase difference in Eq.(\ref{101})
is proportional to
$ \Delta{m}^2_{kj} / 2 E $,
as in
the standard treatment
of neutrino oscillations
in a flat space--time. 

We will now define the phase 
acquired by the mass eigenstate $\nu_k$ when
it travels from point $A(t_A, r_A, \phi_A)$ to point 
$B(t_B, r_B, \phi_B)$ as 
\begin{equation}
\Phi_k = \int_A^B 
\left [E_k dt - p_k(r) dr - J_k d\phi \right ]
\;,
\label{bigphase}
\end{equation}
where we have defined $p_k(r) \equiv - p_r^{(k)}$.
The integration in Eq.(\ref{bigphase}) is performed along the
light--ray trajectory which links the space--time points $A$ and $B$.
At this stage, we note that $E_k$ and $J_k$, which are
constants of motion for the geodesic trajectory of the $k^{\rm th}$
eigenstate, are no longer constant along the light--ray trajectory. Instead,
the energy at infinity $E_0$ and the angular momentum $J_0$ at infinity for
a massless particle are constant along the light--ray path. Therefore,
$E_k$ and $J_k$ cannot be taken out of the integration in
Eq.(\ref{bigphase}) and some caution is required for the calculation.
We will show explicitly in the following Subsections, however, that 
in the relativistic limit, this problem can be circumvented.

We now discuss two different situations: radial propagation 
and non--radial propagation. In the last Subsection we will
address the possibility of gravitational lensing.

\subsection{Radial propagation}

For neutrinos propagating in a radial direction, we have
${\rm d}\phi = 0$ and no angular momentum.
Equation (\ref{bigphase}) is reduced to
\begin{equation}
\Phi_k = \int_{r_A}^{r_B} \left[ E_k 
\left(\frac{dt}{dr}\right)_0 -  p_k(r)\right] dr\;,
\label{phaserad}
\end{equation}
where $p_k(r)$ is obtained, from the mass--shell relation
Eq.(\ref{shell}) with $J_k = 0$, as
\begin{equation}
p_k(r) = \pm \frac{ 1 }{ B(r) }
\,\sqrt{E_k^2-B(r) m_{k}^2}\;,
\label{01}
\end{equation}
and the light--ray differential $(dt/dr)_0$ is
\begin{equation}
\left( \frac{dt}{dr} \right)_0 = \pm \frac{1}{B(r)}\;.
\label{dtdr0-radial}
\end{equation}
In Eqs.(\ref{01}) and (\ref{dtdr0-radial}), 
the sign $(\pm)$ apply to neutrinos propagating outward $(+)$
or inward $(-)$ of the gravitational well, respectively.
Therefore, the quantum mechanical phase $\Phi_k$ is
\begin{equation}
\Phi_{k}
= \pm
\int_{r_A}^{r_B}
\left(
E_{k}
-
\sqrt{
E_k^2
-
B(r)
m_{k}^2
}
\right)
\frac{ dr }{ B(r) }
\;.
\label{05}
\end{equation}
At this point, we apply the relativistic expansion using the energy
at infinity $E_k$ as a reference value, i.e. $ m_k \ll E_k $. 
As in the flat space--time case, the following relation holds
\begin{equation}
E_k
\simeq
E_0
+
{\mathrm{O}}
\left(
\frac{ m_k^2 }{ 2 \, E_0 }
\right)
\;,
\label{06}
\end{equation}
where $E_0$ is the energy at infinity for a massless particle.
Taking into account that
$ 0 < B(r) \leq 1 $,
we have
\begin{equation}
\sqrt{
E_k^2
-
B(r)
m_{k}^2
}
\simeq
E_k
-
B(r)\,
\frac{ m_k^2 }{ 2 \, E_0 }
\;.
\label{07}
\end{equation}
Then, the phase in Eq.(\ref{05}) is approximated by
\begin{equation}
\Phi_{k}
\simeq \pm
\int_{r_A}^{r_B}
\frac{ m_k^2 }{ 2 \, E_0 }\,dr
\;.
\end{equation}
Since the integration is performed along the light--ray trajectory,
$E_0$ is constant and the integration is easily performed to give
\begin{equation}
\Phi_{k}
\simeq 
\frac{ m_k^2 }{ 2 \, E_0 } 
|r_B - r_A|
\label{08}
\;.
\end{equation}
The phase shift which determines the oscillation is, therefore,
\begin{equation}
\Phi_{kj}
\simeq 
\frac{\Delta m_{kj}^2 }{ 2 \, E_0 } 
|r_B - r_A|
\;.
\label{deltaphi}
\end{equation}
We note that the derivation of this result does not 
depend on the weak--field approximation. 

The result for the phase shift in Eq.(\ref{deltaphi}) is in agreement
with that in \cite{BHM}. The authors in \cite{BHM} calculated
the phase of each particle along its classical trajectory and then 
introduced an initial time difference in the phases for the states 
$\nu_k$ and $\nu_j$. Although they arrived at the correct result
(Eq.(\ref{deltaphi})), their derivation is not justified because the
neutrinos produced at different times do not interfere because
they possess initial relative random phases. We believe that the
correct approach is to consider the interference 
between mass eigenstates 
produced at the same space--time position and detected at the 
same space--time point, related by the light--ray relation of 
Eq.(\ref{dtdr0-radial}).
On the other hand, any comparison of our result with that in
\cite{AB} is problematic since the energy $E$ used in
\cite{AB} is not clearly defined among others.

Some comments on the definition of ``relativistic" neutrinos are
in order here. Let us consider the following cases :

\vspace{0.5cm}
\begin{tabular}{llll}
\phantom{space} & (1)~ & $m_k^2 \ll E_k^2$ &
~~: relativistic at infinity,
\\
\phantom{space} & (2)~ & $m_k^2 \ll \left[E^{(loc)}_k(r_A)\right]^2$ &
~~: relativistic at the source,
\\
\phantom{space} & (3)~ & $m_k^2 \ll \left[E^{(loc)}_k(r_B)\right]^2$ &
~~: relativistic at the detector.
\end{tabular}
\vspace{0.5cm}

In case (1), the ratio of $m_k^2$ to any local energy
$E^{(loc)}_k(r)$ is, from Eq.(\ref{eloc})
and $B(r) \leq 1$,
\begin{equation}
\frac{m_k^2}{
	\left[E^{(loc)}_k(r)\right]^2}
=
\frac{m_k^2}{E_k^2}
B(r) 
\;\leq\;
\frac{m_k^2}{E_k^2}
\;\ll\; 1 \;,
\end{equation}
so that the neutrinos are even more relativistic at $r<\infty$, and the
approximation in Eq.(\ref{07}) is certainly justified.

Case (2) needs a caution when the observer happens to be
at infinity, because the ratio of $m_k^2$ 
to the energy at $r=\infty$ becomes
\begin{equation}
\frac
{m^2_k}
{E^2_k}
=
\frac
{m^2_k}
{ \left[E^{(loc)}_k(r_A)\right]^2}
\frac{1}{B(r_A)}\;.
\end{equation}
That is, even if neutrinos are produced highly relativistically,
they are not guaranteed to be relativistic at $r=\infty$,
unless
$ r_A \gg 2GM \left[ 1-(m_k/
E^{(loc)}_k(r_A))^2\right]^{-1}$.
For the obvious reason that
non--relativistic neutrinos cannot be detected,
at least with known techniques
(assuming that neutrino masses are much smaller than 1 MeV),
however,
the lack of validity of
the relativistic condition at infinity
not only means that the approximate formula (\ref{07})
is not valid, but also that in practice such neutrinos
are not detectable at infinity.

Case (3) deals with an observer under the influence of a sizeable 
gravitational field. In this case,
neutrinos stay always relativistic along their path ($r_A < r < r_B$),
which validates approximation in Eq.(\ref{07}), for we have
\begin{equation}
B(r) 
\frac
{m^2_k}
{E^2_k}
=
\frac{B(r)}{B(r_B)}
\frac
{m^2_k}
{ \left[E^{(loc)}_k(r_B)\right]^2}
\;\leq\;
\frac
{m^2_k}
{ \left[E^{(loc)}_k(r_B)\right]^2}
\;\ll\; 1\;.
\end{equation}

In short, neutrinos are assumed to be ``relativistic" when they are
relativistic at infinity, relativistic at the detector,
or relativistic at the production point
with $ r_A \gg 2GM \left[ 1-(m_k/
E^{(loc)}_k(r_A))^2\right]^{-1}$
and then Eq.(\ref{08})
provides the correct quantum phase.

As a final comment, we wish to compare Eq.(\ref{08}) with that
of the flat space--time case. As they stand, the expressions of 
the phase in Eq.(\ref{08}) 
and the phase shift in Eq.(\ref{deltaphi}) appear identical
to those of the flat space--time case.
However, the gravitational effects are present implicitly 
in Eq.(\ref{08}) and Eq.(\ref{deltaphi}). In the absence of a 
gravitational field,
$E_0$ is the energy of the neutrino as seen by any observer along
its trajectory, and $(r_B-r_A)$ is the distance over which a
neutrino propagates. Therefore, Eq.(\ref{08}) gives the standard 
result shown in Eq.(\ref{phase1}). However, in the presence of
gravity, the propagation of a neutrino is over its proper distance
\begin{eqnarray}
L_{p}
\null & \null \equiv \null & \null
\int_{r_A}^{r_B} \sqrt{g_{rr}}\,dr
\label{02}
\\ \nonumber
\null & \null = \null & \null
r_B
\sqrt{ 1 - \frac{ 2GM }{ r_B } }
-
r_A
\sqrt{ 1 - \frac{ 2GM }{ r_A } }
\\ \nonumber
\null & \null & \null
+
2GM
\left[
\ln\!
\left(
\sqrt{ r_B - 2GM }
+
\sqrt{r_B}
\right)
-
\ln\!
\left(
\sqrt{ r_A - 2GM }
+
\sqrt{r_A}
\right)
\right]
\;.
\end{eqnarray}

To simplify the following discussion, 
we consider the case of a weak field,
where $L_p$ is approximated to
\begin{equation}
L_p \simeq r_B - r_A + GM \ln \frac{r_B}{r_A}\;.
\label{eq:r_p}
\end{equation}
This shows that, in a gravitational field, the effective length
in the phase (i.e. $(r_B-r_A)$)
is shorter than $L_p$.
Moreover, the energy measured by a 
detector at $r_B$ is not $E_0$, but rather the
local value $E^{(loc)} (r_B)$. When expressed in terms of the
local energy and the proper distance, the phase shift $\Phi_{kj}$ of 
Eq.({\ref{deltaphi}) in the weak field approximation is
\begin{equation}
\Phi_{kj} \simeq 
\left(
\frac{ \Delta m^2_{kj} L_p}{2 E^{(loc)}(r_B)}
\right)
\left[
1
-
G M
\left(
\frac{1}{L_p}
\,
\ln \frac{r_B}{r_A}
-
\frac{1}{r_B}
\right)
\right]
\;.
\label{phi_measured}
\end{equation}
The first parenthesis on the right--hand side in Eq.(\ref{phi_measured})
is analogous to the flat space--time oscillation phase. The second
parenthesis represents the correction due to the gravitational effects.

The proper oscillation length $L_{kj}^{\mathrm{osc}}$, in the weak
field approximation, is
\begin{equation}
L_{kj}^{\mathrm{osc}}
(r_B)
=
\frac{ 4 \pi E^{\mathrm{loc}}(r_B) }{ \Delta{m}^2_{kj} }
-
G M
\left[
\ln\!\left(
1
-
\frac{ 4 \pi E^{\mathrm{loc}}(r_B) }{ \Delta{m}^2 \, r_B }
\right)
+
\frac{ 4 \pi E^{\mathrm{loc}}(r_B) }{ \Delta{m}^2 \, r_B }
\right]
\;,
\end{equation}
where the quantity in square brackets is negative.
We conclude, therefore, that the proper oscillation length is 
increased in the gravitational field, as expected.

\subsection{Non--radial propagation}
   
In this Subsection, 
we discuss the case of the propagation along a general
trajectory. 
In contrast to the radial case, the motion has an
additional angular dependence. 
The phase $\Phi_k$ is
\begin{equation}
\Phi_k =
\int_{r_A}^{r_B}
\left[
E_k 
	\left(
\frac{dt}{dr}
	\right)_0
- p_k(r)
-J_k
	\left(
\frac{d\phi}{dr}
	\right)_0
\right]
dr\;,
\label{eq:PhiD}
\end{equation}
where the integral is taken
along the light--ray trajectory that links the production point $A$ 
to the detection point $B$. In Eq.(\ref{eq:PhiD}), the quantities
$(dt/dr)_0$ and $(d\phi/dr)_0$ along the light--ray trajectory are
\begin{eqnarray}
	\left(
\frac{dt}{dr}
	\right)_0
&=&
\frac{E_0}{B^2(r) p_0(r)}\;,
\label{l_ray}
\\ \nonumber
	\left(
\frac{d\phi}{dr}
	\right)_0
&=&
\frac{J_0}{r^2}
\frac{1}{B(r) p_0(r)}
\;.
\end{eqnarray}
It is convenient to express the angular momentum $J_k$
as a function of the energy $E_k$, the impact parameter $b$ and
the velocity at infinity $v_k^{(\infty)}$ \cite{Weinberg}
\begin{equation}
J_k = E_k\,b\,v_k^{(\infty)}\;.
\label{jk}
\end{equation}
Since at $r=\infty$
the metric is Minkowskian (no gravity), we can
write
\begin{equation}
v_k^{(\infty)} = \frac{\sqrt{E_k^2 - m_k^2}}{E_k} \simeq
1 - \frac{m_k^2}{2 E_k^2}\;,
\label{v_k}
\end{equation}
where in the last equality we used the relativistic approximation
up to the order ${\mathrm{O}} (m_k^2/E_k^2)$. 
The angular momentum of a massless particle,
$J_0$, is obviously
\begin{equation}
J_0 = E_0\,b\;.
\label{j0}
\end{equation}
With Eqs.(\ref{l_ray})--(\ref{j0}), the expression of $\Phi_k$ in
Eq.(\ref{eq:PhiD}) can be conveniently arranged as follows
\begin{equation}
\Phi_k = 
\int_{r_A}^{r_B} 
dr 
	\frac{E_0}{B(r) p_0(r)}
	\left[
	\frac{E_k}{B(r)}
	-\frac{B(r) p_0(r)}{E_0} p_k(r) 
	- \frac{E_k b^2}{r^2}
		\left(
		1 - \frac{m_k^2}{2 E_k^2}
		\right)
	\right]\;.
\label{46}
\end{equation}
The mass--shell condition Eq.(\ref{shell}) gives
\begin{eqnarray}
B(r) p_0(r) &=& 
\pm
E_0
\sqrt{
	1-B(r) \frac{b^2}{r^2} 
     }\;,
\label{bpr0}
\\ 
B(r) p_k(r) &=&
\pm
E_k \sqrt{1-B(r) \frac{b^2}{r^2} - B(r) \frac{m_k^2}{E_k^2} 
\left ( 1 - \frac{b^2}{r^2} \right)}
\label{bpr}
\\ \nonumber
&\simeq& \pm
E_k \sqrt{1-B(r) \frac{b^2}{r^2} }
\left[
1-\frac{B(r)(1-b^2/r^2)
       }{
        1-B(r) b^2/r^2
       }
\frac{m_k^2}{2E_k^2}
\right]\;.
\end{eqnarray}
The last approximate equality in Eq.(\ref{bpr}) is due to the
relativistic expansion. In Eqs.(\ref{bpr0}) and (\ref{bpr}),
the sign $(\pm)$ is determined by whether $dr$
is positive ($+$) or negative ($-$). 
Substitution of Eqs.(\ref{bpr0}) and (\ref{bpr}) into Eq.(\ref{46})
simplifies
the expression for the phase $\Phi_k$ to
\begin{equation}
\Phi_k \simeq \int_{r_A}^{r_B} dr 
\frac{E_0}{B(r) p_0(r)} E_k \frac{m_k^2}{2 E_k^2}
\;.
\end{equation}
Since the following relation holds,
in the relativistic approximation of Eq.(\ref{06}),
\begin{equation}
E_k \frac{m_k^2}{2 E_k^2} \simeq E_0 \frac{m_k^2}{2 E_0^2}
\;,
\end{equation}
$\Phi_k$ can be expressed as
\begin{equation}
\Phi_k \simeq \pm\frac{m_k^2}{2E_0}
\int_{r_A}^{r_B}
\frac{dr}{\sqrt{
		1-B(r) (b^2/r^2)
	       } }\;.
\label{phase-nonr}
\end{equation}

Equation (\ref{phase-nonr}) is the phase acquired by
the mass eigenstate $|\nu_k\rangle$ for a non--radial propagation
from the source $A$ to the detector $B$. 
In the limit $b\rightarrow 0$, which reduces the motion to be radial,
Eq.(\ref{08}) is recovered. We also notice that the integrand in
Eq.(\ref{phase-nonr}) is divergent at the point of the closest approach
$r_0$, defined by the condition that the rate of change of the
coordinate $r$ with respect to the angle $\phi$ vanishes:
\begin{equation}
\frac{dr}{d\phi} = 0 \;\;\; \Longrightarrow \;\;\;
E_0^2 = \frac{J_0^2}{r_0^2} B(r_0)
\;\;\; \Longrightarrow \;\;\;
1 - \frac{b^2}{r_0^2} B(r_0) = 0\;.
\label{closest}
\end{equation}
However, the integral which gives the phase $\Phi_k$ is finite.
We will show this explicitly in the weak field approximation.

The expression of $\Phi_k$ obtained in Eq.(\ref{phase-nonr}) is valid 
for any spherically symmetric (and time--independent) field. 
It has been derived without 
any assumption on the strength of
the gravitational field. 
In order to gain more physical 
insight, however, we perform a weak field approximation, which allows us to
perform the integration analytically. The approximation is valid
if the field is weak enough to satisfy the condition $GM \ll r$
for all the $r$'s along the trajectory under consideration. 
For example, the gravitational field of the sun at its surface
is about $GM_\odot/R_\odot \sim 2 \times 10^{-6}$
and that of a galaxy is about
$G(10^{11}M_\odot)/30 \mbox{kpc} \sim 1.6 \times 10^{-7}$, 
both of which justify the weak field approximation.
Whenever the weak field approximation is applied, we
keep the expansion up to the order ${\mathrm{O}}(GM/r)$.

First, let us consider the case where
a neutrino is produced in a gravitational field and then 
propagates outward from the potential well non--radially. 
The weak field approximation allows us to expand
\begin{equation}
\sqrt{
1-B(r) \frac{b^2}{r^2} 
}
\simeq
\sqrt{
1-\frac{b^2}{r^2}
}
\left[
1 + \frac{GM}{r}
\frac{b^2}{r^2-b^2}
\right]\;.
\label{eq:expand1}
\end{equation}
The phase $\Phi_k$ is then easily integrated and becomes
\begin{equation}
\Phi_k \simeq 
\frac{m_k^2}{2E_0}
\left[
\sqrt{r_B^2 - b^2}
-
\sqrt{r_A^2 - b^2}
+  GM
\left(
\frac{r_B}{ \sqrt{r_B^2 - b^2}}
-
\frac{r_A}{ \sqrt{r_A^2 - b^2}}
\right)
\right]\;.
\end{equation}
We notice again, as a consistency check, that the radial limit
$b\rightarrow 0$ gives the same expression as given in Eq.(\ref{08}).
  
The second situation is when a neutrino moves around 
the massive object,
crossing the closest approach point at $r=r_0$.
Taking into account the sign of the momentum, the phase is
\begin{equation}
\Phi_k(r_A \to r_0 \to r_B) = 
\frac{m_k^2}{2E_0}
\int_{r_0}^{r_A} 
\frac{dr}{\sqrt{
		1-B(r) (b^2/r^2)
	       } }
+
\frac{m_k^2}{2E_0}
\int_{r_0}^{r_B} 
\frac{dr}{\sqrt{
		1-B(r) (b^2/r^2)
	       } }\;.
\label{phi-case-2}
\end{equation}
The position of the closest approach can be
solved from 
Eq.(\ref{closest}) in the weak field approximation,  as
\begin{equation}
r_0 = b \left( 1- \frac{GM}{b} \right)\;.
\label{58}
\end{equation}
Substituting Eq.(\ref{58}) into Eq.(\ref{phi-case-2}), we have
\begin{eqnarray}
\Phi_k &\simeq& 
\frac{m_k^2}{2E_0}
\left[
\sqrt{r_A^2 -r_0^2}
+
\sqrt{r_B^2 -r_0^2}
+
GM
	\left(
	\sqrt{
		\frac{r_A -r_0} {r_A +r_0}
	      }
	+
	\sqrt{
		\frac{r_B -r_0} {r_B +r_0}
	      }
	\right)
\right]
\label{phiturn}
\\ 
&\simeq& 
\frac{m_k^2}{2E_0}
\left[
\sqrt{r_A^2 -b^2}
+
\sqrt{r_B^2 -b^2}
\phantom{\frac{GM b}{\sqrt{r_A^2 -b^2}}}
\right.
\nonumber 
\\
& & 
\left.
+ GM \left( \frac{b}{\sqrt{r_A^2 -b^2}} +
              \frac{b}{\sqrt{r_B^2 -b^2}} +
              \sqrt{\frac{r_A -b}{r_A +b}} +
              \sqrt{\frac{r_B -b}{r_B +b}}
       \right)
\right]
\nonumber \;.
\end{eqnarray}
We observe that, in this case, the radial limit $b\rightarrow 0$
is meaningless, because it would correspond to a radial motion
which crosses the gravitational source, where our description
becomes inadequate.

For $b \ll r_{A,B}$, 
Eq.(\ref{phiturn}) is reduced to
(up to the order of $(b^2/r_{A,B}^2)$)
\begin{equation}
\Phi_k= 
\frac{m_k^2}{2E_0}
(r_A + r_B)
\left[
1 - \frac{b^2}{2 r_A r_B}
+ \frac{2GM}{r_A + r_B}
\right]\;,
\label{eq:FinalPhi}
\end{equation}
which will be used to discuss the gravitational
lensing in the following Subsection.
It is interesting to note that 
Eq.(\ref{eq:FinalPhi}) has a gravitational effect which
does not depend on the distance
between the source and the detector (assuming that this distance is
much larger than the impact parameter $b$).
That is, this gravitational effect integrated along a trajectory
which passes close to a gravitational center
induces a constant phase shift $ 2GM (m_k^2/2E_0) $.
Furthermore, this constant phase shift
does not depend on how close the
trajectory passes to the gravitational center.
Therefore, if, for example, a neutrino travels relatively close
to several well--separated gravitational centers,
the net phase shift becomes
the sum of the phase shifts induced by each gravitational center.

\subsection{Gravitational Lensing of Neutrinos}

Let us consider a gravitational lens which is located
between a source and an observer but off the line connecting 
the two. A neutrino emitted from the source can travel
along two different paths, the proper distances of which
are different
and give the quantum interference at 
the detector. 

Oscillations arise due to the interference not only between 
the mass eigenstates
$\nu_k$ and $\nu_j$ travelling along
each path, but also between the mass eigenstates propagating
along different paths (for definiteness, we denote them as
long--path ($L$) and short--path ($S$) ).
A neutrino produced as a flavor eigenstate
$ |\nu_e\rangle =\cos \theta |\nu_1\rangle 
+ \sin \theta |\nu_2\rangle$
at the source $A(t_A, r_A, \phi_A)$,
evolves into
(we consider only two generations)
\begin{equation}
| \nu_e, B \rangle = N \sum_{path=L,S}
\left[
\cos\theta  \exp{(-i \Phi_1^{path}) }|\nu_1\rangle
+
\sin\theta \exp{(-i \Phi_2^{path}) } |\nu_2\rangle
\right]
\;,
\label{prob0}
\end{equation}
where $N$ is the normalization constant. 
The flavor--changing oscillation probability at the detector
is then given by
\begin{eqnarray}
{\cal P\/}(\nu_e \to \nu_\mu)
&=&
| \langle \nu_\mu| \nu_e, B \rangle |^2
\label{eq:prob}
\\ \nonumber
&=&
\frac{1}{2}\cos^2\theta \sin^2\theta
\left[
1+ 
\cos(\Phi_1^L-\Phi_1^S)
+1+ \cos(\Phi_2^L-\Phi_2^S)
\right.
\\ \nonumber
& &
\hspace{0.7cm}
 -\left\{
\cos(\Phi_2^L-\Phi_1^L)
+
\cos(\Phi_2^S-\Phi_1^S)
\right\}
\\ \nonumber
& &
\hspace{0.7cm}
\left.
 -\left\{
\cos(\Phi_2^L-\Phi_1^S)
+
\cos(\Phi_2^S-\Phi_1^L)
\right\}
\right]
~.
\end{eqnarray}
The phases $\Phi_k^{path}$ in Eq.(\ref{eq:prob})
can be evaluated along the light--ray trajectories
as shown in the previous Subsection.
Substituting Eq.(\ref{eq:FinalPhi}) into Eq.(\ref{eq:prob}),
we have
\begin{eqnarray}
{\cal P\/}(\nu_e \to \nu_\mu)
&=& \sin^2(2\theta) 
\left[
\sin^2 \left\{
	\frac{\Delta m^2X}{4E_0}
		\left(
		1 + \frac{2GM}{X} - 
			\frac{\Sigma b^2}{ 4 r_A r_B}
		\right)
	\right\}	
\right.
\label{finalfinal}
\\ \nonumber
& &
\hspace{0.7cm}
\times
\cos	\left(
	\frac{m^2_1X}{4E_0}
			\frac{\Delta b^2}{ 2 r_A r_B}
	\right)
\cos	\left(
	\frac{ m^2_2 X}{4E_0}
			\frac{\Delta b^2}{ 2 r_A r_B}
	\right)
\\ \nonumber
& &
\hspace{0.7cm}
\left.
+
\sin^2 \left(
	\frac{\Sigma m^2 X}{4E_0}
	\frac{\Delta b^2}{ 4 r_A r_B}
       \right)
\sin^2 \left(
	\frac{\Delta m^2 X}{4E_0}
	\frac{\Delta b^2}{ 4 r_A r_B}
       \right)
\right]
~,
\end{eqnarray}
where we have defined
$X \equiv r_A + r_B$,
$\Delta m^2 \equiv m_2^2 -m_1^2 $,
$\Sigma m^2 \equiv m_2^2 +m_1^2 $,
$\Delta b^2 \equiv b_L^2 -b_S^2 $ and
$\Sigma b^2 \equiv b_L^2 +b_S^2 $.
In the symmetric case where the lens is aligned with the
source and the detector, $\Delta b^2=0$ and $\Sigma b^2 = 2b^2$ and
the above flavor--changing probability is reduced to that of the
non--radially propagating neutrinos
\begin{equation}
\left. {\cal P\/}(\nu_e \to \nu_\mu)\right|_{(\Delta b^2=0)}
= \sin^2(2\theta) 
\sin^2 \left[
	\frac{\Delta m^2X}{4E_0}
		\left(
		1 + \frac{2GM}{X} - 
			\frac{ b^2}{ 2 r_A r_B}
		\right)
	\right]
~,
\end{equation}
which can be obtained directly from Eq.(\ref{eq:FinalPhi}). This
is expected, since the symmetric case is equivalent to 
the case of the non--radial propagation.

Obviously, the proper way to discuss the gravitational lensing 
effects on the neutrino oscillations would require a wave packet
formalism.
Such a study is beyond the scope of the present paper and
will be given elsewhere.

\section{Conclusion}
\label{sec:final}

We have studied the propagation of neutrinos in a curved space--time
and the modification to the neutrino oscillation
by calculating a covariant quantum mechanical phase
$\Phi_k$.
The gravitational field considered in this work
is that of a non--rotating spherically 
symmetric object, described by the Schwarzschild metric.
Furthermore, we have assumed that neutrinos 
are relativistic so that a plane wave 
analysis can be applied.

Radial and non--radial propagation  have
been discussed in the light--ray
approximation.
Although our phase for the radial motion is in agreement with the result
of the previous work\cite{BHM},
the interpretations are different.
Any comparison of our result with that in
\cite{AB} is problematic since the energy used in
\cite{AB} is not clearly defined.

The calculated phase appears identical to 
that of the flat space--time case.
This is because the phase is expressed in terms of the
asymptotic energy $E$ and the coordinate distance.
However, the gravitational effects do appear in the leading order if
we express the phase with
the locally measured energy
and proper distance.
As in the radial case, the phase of 
relativistic neutrinos for the non--radial motion
has been obtained without resort to the weak field
approximation.
Assuming that the gravitational field is
weak enough and the source and the detector are at a 
sufficiently large distance
from the massive object,
the phase is reduced to a simpler form 
as given in Eq.(\ref{eq:FinalPhi}).
Finally, we have considered the gravitational lensing of neutrinos,
i.e. the quantum interference when neutrinos propagate through
different paths, and have derived the flavor--changing probability
${\cal P\/} (\nu_e \to \nu_\mu)$ as given in Eq.(\ref{finalfinal}).

Even though the measurement of the gravitational effects on the 
propagation and oscillations of neutrinos 
is not feasible at present,
we think that the understanding of these behaviors themselves is
of interest.

\vspace{1cm}
{\bf Acknowledgements.}
NF gratefully acknowledges a fellowship from the Istituto Nazionale
di Fisica Nucleare, Italy.

\vspace{1cm}
{\bf Note added.} 
After the completion of this paper, 
we became aware of the paper by 
C.Y. Cardall and G.M. Fuller \cite{Cardall},
which discussed a similar subject and obtained the results
for the radial motion similar to ours.

\end{document}